\documentclass[twocolumn,aps,showpacs]{revtex4}
\usepackage[dvips]{graphicx,color}
\newcommand{\p}{\partial}
\newcommand{\ath}{\alpha}
\newcommand{\af}{\tilde{\alpha}}
\newcommand{\ad}{\xi}
\newcommand{\lp}{\left(}
\newcommand{\rp}{\right)}
\newcommand{\lf}{\left [}
\newcommand{\rf}{\right ]}
\newcommand{\be}{\begin{equation}}
\newcommand{\ee}{\end{equation}}
\newcommand{\erf}{\mbox{erf}}
\newcommand{\eps}{\varepsilon}
\newcommand{\mean}[1]{\left\langle {#1} \right\rangle}
\newcommand{\meanet}[3]{\left\langle {#1}\left|{#2}\right|{#3}\right\rangle}
\newcommand{\mycomments}[1]{}
\newcommand{\ka}{\kappa}
\newcommand{\kap}{\mathcal{K}}

\newcommand{\mys}{s}

\newcommand{\hl}{\hat{L}}
\newcommand{\hld}{\hat{L}^{\dagger}}
\newcommand{\vecr}{\vec{r}}
\newcommand{\vecR}{\vec{R}}
\newcommand{\vecq}{\vec{q}}
\newcommand{\vell}{\vec{\ell}}

\begin{document}
\title{Stability and Dynamics of Crystals and Glasses of Motorized Particles}
\author{Tongye Shen and Peter G. Wolynes}
\affiliation{Department of Chemistry \& Biochemistry, Department of Physics,
University of California, San Diego and Center for Theoretical Biological
Physics, La Jolla, California, 92093-0371}
\date{\today}
\begin{abstract}

Many of the large structures of the cell, such as the cytoskeleton,
are assembled and maintained far from equilibrium. We
study the stabilities of various structures for 
a simple model of such a far-from-equilibrium
organized assembly in which spherical particles
move under the influence of attached motors. 
From the variational solutions of the manybody master equation
for Brownian motion with motorized kicking we obtain
a closed equation for the order parameter of 
localization. Thus we obtain the transition criterion
for localization 
and stability limits for the crystalline phase and frozen amorphous
structures of motorized particles. 
 The theory also allows
an estimate of nonequilibrium effective temperatures characterizing
the response and fluctuations of motorized crystals and
glasses.

\end{abstract}
\maketitle

Assemblies of molecular sized particles are seldom
far from equilibrium owing to the relative strength
of the thermal buffeting inherent at this scale.
As we consider assemblies of larger and larger
particles the thermal forces become less capable
of moving and reorganizing such assemblies.
At the size scale of biological cells, objects
are not rearranged just by equilibrium thermal
forces but are moved about by motors or by polymerization
processes that use and dissipate chemical energy~\cite{Pollard}.
What are the rules that govern the formation
of periodically ordered or permanently organized 
assemblies at this scale? Does the far-from-equilibrium 
character of the fluctuating forces
due to motors and polymer assembly change the
relative stability of different colloidal
phases?  These problems are not unique 
for intracellular dynamics, but belong to an emerging 
family of nonequilibrium assembly
problems ranging from driven 
particles~\cite{VCBCS95}, swarms~\cite{EE03}, and jamming~\cite{LN01,TPCSW01}, 
to microscopic pattern formation 
and mesoscopic self-organization~\cite{KGWW03}.

Motivated by these considerations which may be relevant
for the dynamics of the 
cytoskeleton~\cite{Pollard,Amos91}  as well as other
far-from-equilibrium aggregation systems, we study 
a simple motorized version of the standard hard sphere
fluid often used to model colloids.
Both motors and nonequilibrium polymer assemblies
can convert the chemical energy of high energy
phosphate hydrolysis to mechanical motions which
one would ordinarily think would ``stir'' and hence
destabilize ordered structures.
We will show these systems in some circumstances may have an enlarged
range of stability relative to those with purely thermal motions.

We adopt a stochastic description of the motions of a collection of 
motorized particles. The overdamped  Langevin dynamics is
$\dot{\vecr_i} = \beta D \vec{f}_i +  \vec{\eta}(t)+ \vec{v}^m_i$.
Here $\vecr_i$ is the position of the $i$th particle, 
$\vec{f}_i =-\nabla_i U$ is the mechanical force that 
comes from the usual potential $U(\{\vecr\}):=U(\vecr_1, \vecr_2,\ldots, \vecr_n)
=\sum_{\mean{ij}} u(\vecr_{ij})$ 
among particles. 
The random variable $\vec{\eta}$ vanishes on average and is Gaussian with
$\mean{{\eta}^\alpha_i(t) {\eta}^\beta_j(t')} = 2 D \delta_{\alpha\beta}
\delta_{ij}\delta(t-t').$  
The motor term $\vec{v}^m(t)= \sum_q \vec{\ell}_q \delta(t-t_q)$
is a time series of shot-noise-like kicks. Its properties depend
on the underlying biochemical mechanism of the motors.
The stochastic nature of the motors also leads to a
master equation description~\cite{Kampen92,Risken}  for the dynamics 
of the probability distribution function $\Psi$ 
of the particle configurations,
\be
\left[ {\p\over \p t} - (\hl_{FP}+\hl_{NE})\right]\Psi(\{\vecr\},t) = 0
\ee
Here $\hl_{FP}:=D \sum_i \nabla_i\cdot(\nabla_i
-  \beta \vec{f}_i )$ is the Fokker-Planck operator.
An integral operator 
$\hl_{NE}\Psi(\{\vec{r}\}) = \int \Pi_i d\vec{r'}_i\lf K_{\{\vec{r'}\}\to \{\vecr\}} \Psi(\{\vec{r'}\})
-K_{\{\vecr\}\to \{\vec{r'}\}} \Psi(\{\vecr\})\rf$ summarizes the nonequilibrium kicking
effect of the motors. 

The motors are firmly built in the particles. They work by
consuming chemical energy sources, like ATP. In a single chemical
reaction event, the motor makes a power stroke (which induces a discrete conformation
change) that moves the particle by a distance of $\ell$ in the direction $\hat{n}$.
Motor kicking can be modeled as a two-step stochastic process: 
step I, the energy source binds to the motor, and the 
step II, 
the reaction ensues and the resulting conformational change  makes  
a power stroke. The rate of the first step $k_1$ depends on the energy source
concentration while the rate of the second step $k_2$ depends on the
coupling between the structural rearrangement and the external forces, 
 $k_2= \ka\exp( \mys \beta [U(\vecr)-U(\vecr+\vec{\ell})])$, i.e.,
motors slow down when 
they work against mechanical obstacles. Such slowing has been demonstrated
in the case of microtubules~\cite{DY97,Howard}.
$\mys$, the coupling strength, measures the relative location of the transition
state for the power stroke step and ranges from 0 to 1.
At the limit $s \to 1$ we have a {\em susceptible} motor and while $\mys \to 0$ 
corresponds to
an {\em adamant} motor. We use these names in the sense that an adamant motor
is not sensitive to its thermal-mechanical environment, so each power stroke uses and wastes
a lot of energy; in contrast a susceptible motor saves energy 
running faster downhill (free energy) and slower uphill. 

 We assumed that step II is the bottleneck of kinetics, i.e., the overall rate 
$k\approx k_2\ll k_1$.
To make our model
suitable for a variety of situations, we specify different variables $s$ and $s'$ for
the degree of susceptibility for going uphill and downhill respectively.
 Thus $k= \ka \exp(-\beta \mathcal{G}[U(\vecr+\vec{\ell})-U(\vecr)])$ with  
 a switch function $\mathcal{G}(x):= \Theta(x) s x + \Theta(-x) s' x$.
 Here $\Theta$ is the Heaviside function.
 Thus the overall temporal statistics of the kicks is position-dependent Poisson
distribution.

The kicking direction $\hat{n}(t)$ fluctuates on the 
time scale $\tau$ of the particle tumbling.
We will study below explicitly two extremes:
the {\em isotropic } kicking case when $\tau$ is very small and
the {\em persistent} kicking case
when $\tau$ is very large compared with $\ka^{-1}$, i.e., 
each motor always kicks in a predefined direction. The
direction of persistence will be assumed to vary randomly from particle
to particle.

To solve the dynamics of probability distributions,
researchers often pose the problem as the solution of 
a variational problem. 
 Due to the $\hl_{NE}$ part of eqn.~(1), we can
not  perform the usual transformations of the left and 
right state vector to make $\hl$ hermitian~\cite{Risken}.
For this type of problems, nevertheless we can 
 obtain the solution of the many-particle
master equation using nonhermitian variational
methods as described by Eyink~\cite{Eyink96,EA97} or
using 
the squared (therefore hermitianized) operator $\hld\hl$ (e.g., p159 of~\cite{Risken}).

Eyink's nonhermitian variational formulation
is similar to the  Rayleigh-Ritz method in ordinary quantum
mechanics but uses independent left and right state vectors. 
For isotropic kicking, we start with a Jastrow-like trial function
\be
\Psi(\{\vecr\})= 
\exp\left\{ - \sum_i [\ad_i (\vecr_i -\vec{R}_i)^2] - \beta  U(\{\vecr\})\right\}
\ee
Similar to the quantum hard spheres~\cite{J55},
such a trial function avoids any singularities of  $\hl_{FP}$ arising from
the hardsphere potentials $u_{ij}(r_{ij})$ between particles $i$ and $j$ used
in this study.
For simplicity, we set $\vecq_i:=\vecr_i-\vec{R}_i$ and a uniform $\ad_i:=\ad$.
The nonhermitian variational method implies
for steady states that the second moment of $\vecq$ satisfies
the moment closure~\cite{CCK89} equation
\be
{\p \mean{q_j^2}\over \p t}=\mean{q_j^2 \left|\hl_{FP}+ \hl_{NE} \right| \Psi}=0
\ee

To effectively evaluate eqn.~(3), we need to simplify the
many-body integration $\int \Pi_i d^3 \vecq_i $ involving
$\exp-\beta U(\{\vecr\})$. Here we use cluster expansion~\cite{Fixman69,SW84} 
to render the manybody Boltzmann factor a product of effective 
single body terms by averaging over the neighbors' fluctuations. 
We thus have
 $e^{-\beta U}=\Pi_i e^{-\beta u_i} \approx \Pi_i e^{-\beta \hat{u}_i}$. 
Here the original $u_i= \sum_{j\ne i}{1\over 2} u_{ij}$ depends on
the manybody configuration, while $\hat{u}_i$ depends on $\vecq_i$ (and
constant $\{\vecR\}$) only. In fact, we keep it to the harmonic 
order for consistency, i.e., $\hat{u}_i= \beta^{-1}\ath q_i^2$. Here $\ath$ is the effective
spring constant from the mechanical feedback from neighbors. $\ath$ depends
on its neighbors' overall fluctuations controlled by
$\af$ and their mean position $\{\vec{R}\}$. In turn, the positions of the neighbors 
are controlled by the lattice spacing
for crystals or radial distribution functions for glasses and ultimately
by the nature of the structure and the particle density $n$.
I.e., for fcc lattice, we have $\ath_{cr}=\ath_{cr}(\af, n)$ as
the eigenvalues of the Hessian matrix constructed from the effective
potential ${1\over 2}\sum_{j\in n.n.} v(|\vecR_j|;\af)$. 
Here $v(R;\af)= \ln\{1+ {1\over 2}\erf[{(R-1)\sqrt{\af}}] - {1\over 2}\erf[{(R+1)\sqrt{\af}}]
+{(\af\pi)^{-1/2}\over 2R} [e^{-\af (R-1)^2} - e^{-\af (R+1)^2}]\}$
and  the sum of $\vecR_j$ is over the 12 nearest neighbor
positions of the origin of a fcc lattice (with lattice spacing $({4\over n})^{1\over 3}$). 
For glasses~\cite{SSW85,HW03}, we replace
the summation over discrete crystalline neighbor location
with a mean-field average over the first shell of the
pair distribution function of the hardsphere liquids,
$\ath_{gl}(\af, n)= {n\over 6}\int_{1st.sh.} g(R,n)~Tr \nabla\nabla v(\vec{R};\af)d\vec{R}$.
For numerical work, we take $g(R,n)$ as the Verlet and Weis's corrected
radial distribution function~\cite{VW72}.

After some calculations, we obtain the steady-state
manybody probability distribution function as a product of
localized Gaussians of the form $\exp[-\af (\vecr_i -\vec{R}_i)^2]$
with $\af$, the final localization strength, satisfying two equations: 
\be
\af= \ad + \ath(\af, n)
\ee
\be
6D {\ad \over\af}+ \ka I_2(\af)\times ({\pi\over \af})^{-3/2} =0
\ee
The first and second terms of eqn.~(5) come from $\meanet{q_j^2}{\hl_{FP}}{\Psi}\over
\mean{1 | \Psi}$ and $\meanet{q_j^2}{\hl_{NE}}{\Psi}\over
\mean{1 | \Psi}$ respectively.
The integral $I_{n}(\af):=\int d^3\vecq [(\vecq+\vell)^n-\vecq^n]e^{-\af q^2+
\mathcal{G}( -2\af \vell\cdot\vecq -\af \ell^2
)}$ can be further expressed as explicit but complicated analytical formulas with
incomplete Gamma functions. 
Thus using eqn.~(4) in eqn.~(5), we finally derive the
order parameter for localization, $\af$, in a closed form
with parameters $\ell, D, \ka, s, s'$ and $n$.
When $\ka\cdot\ell=0$, we have $\ad=0$ and 
the equilibrium equation $\af=\ath(\af,n)$, which returns
to the self-consistent phonon solution~\cite{Fixman69,SW84,SSW85,MP99,HW03}.
A nonzero solution $\af$ of eqn.~(5) is only obtained at sufficiently high density, i.e.,
for $n>n_c$. 
For low density, the system cannot support stable localized vibrations and is
in the fluid phase with $\af=0$. An instability density $n_c$ separates these two phases.
This phase transition is first-order like, characterized by a discontinuous jump of $\af$.

We calculated $n_c$ as a function of two independent parameters $s$ and $s'$, $n_c(s,s')$
for various $\ka$, $D$, and $\ell$.
An important dimensionless
ratio $\Delta:=\ka\ell^2/D$ measures the strength of chemical {vs.}~thermal noise.
For an actin polymer solution~\cite{Pollard}, we can relate
the effective
kicking rate to the speed of nonequilibrium polymerization.
Here $\ell$ is the monomer size $0.01\mu m $. ($\mu=10^{-6}$) The
treadmill concentration is
$c_{tr}=0.17 {\mu}M$.
The chemical reaction rates of the barbed and pointed ends of actin are
$k_+=11.6{\mu}M^{-1}s^{-1}$ and $k_-= 1.3{\mu}M^{-1}s^{-1}$ respectively.
The diffusion constants of rod are given by the Kirkwood equation~\cite{DoiEdwards}.
This equation relates the diffusion to solvent viscosity $\eta_s$ and gives
the translational diffusion constants: $D_{\|} = k_B T \ln( L/d)/ (2 \pi \eta_s L)$ and
$D_{\bot}= D_{\|} /2$. Using the typical length of the actin
filaments  $L\sim 20{\mu}m$ and typical width $d$ of about $0.015\mu m$.
An estimate of the hydrodynamic diffusion along the rod gives $D_\|\sim 0.1 {\mu}m^2 s^{-1}$.
Thus for a dilute solution of typical actin filaments, $\Delta\sim 10^{-3}$-$10^{-2}$. However,
in vivo, the actin monomer concentration is kept much higher than $c_{tr}$ (with the help of
capping proteins which prevent actin filaments from growing longer).
Also the viscosity of the cell medium is higher than $\eta_s$ of pure water due to the presence of
other macromolecular components.
These lower the effective diffusion constant and
raise the value of $\ka$, and therefore they could push $\Delta$ above 1.
The limit $\Delta\gg 1$ corresponds to entirely motorized motion.

The resulting densities $n_c(s,s')$ are shown in Fig.~1 and 2.
Figure~1 shows some 1D plots that come from 
vertical slices of $n_c$  for several special cases.
Figure~2 shows the 2D top-view of the critical surfaces $n_c$ for
variety of parameters. For the case shown with $\Delta \gg 1$,
$\ath(n)$ (which is the only $n$ dependent part of $\af$) drops to zero, 
while we still have a nonzero solution $\af=\ad$ 
for a corner of $(s,s')$ space. 
In the opposite corner 
we stopped searching for solutions when $n_c$ approached
the maximum packing density $\sqrt{2}$. Thus we have
three distinct regions. 

As seen from these figures,
kicking noise does not always destabilize the structures.
Instead the localized phases have an enlarged
stability range when $s+s'> 1$. When $s'=1-s$, the
same stability limits are obtained as in the equilibrium thermodynamic theory.
Both the frozen disordered glass and the ordered
fcc lattice can be stabilized by kicking motors. The motor effects on the
fcc lattice are more pronounced. The fcc phase has
a larger stable region than the glass.

\begin{figure}
\includegraphics[scale=0.5, angle=-90]{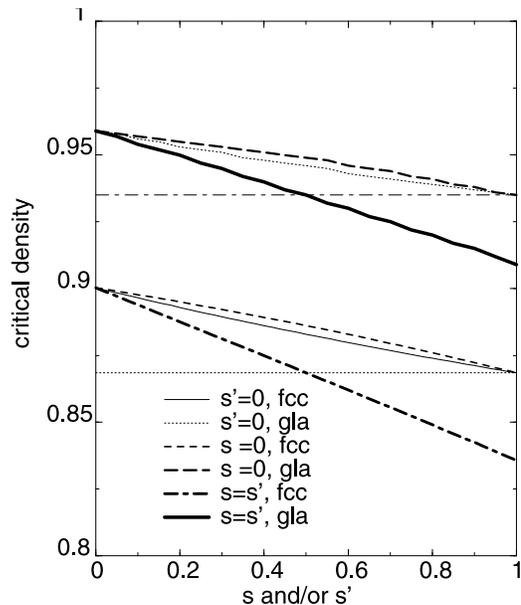}
\caption{ The instability density of the motorized fcc lattice
and the glass as functions of the coupling parameters for
these cases 
i) $s=0$ ii) $s'=0$  and iii) $s =s'$.
In these plots there is 
3D isotropic kicking with $D=0.1$, $\kappa=10$, and $\ell=0.05$.
Therefore $\Delta={\ka \ell^2\over D}=0.25$. 
The two horizontal lines are the corresponding equilibrium
 ($\kappa \ell=0$) cases. 
\label{fig1}}
\end{figure}

\begin{figure}
\includegraphics[scale=0.5, angle=0]{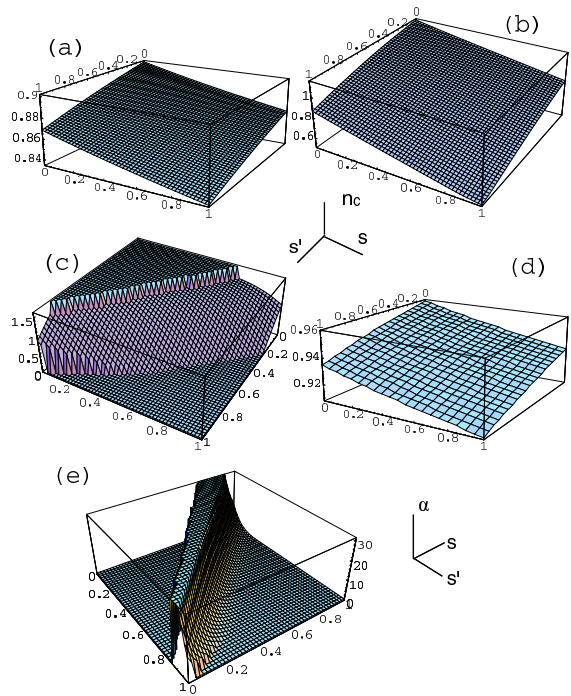}
\caption{ The instability densities $n_c(s, s')$ surface are shown 
for motorized fcc lattice cases: (a) $\Delta=0.25$,
(b) $\Delta=2.5$, (c), $\Delta=25$ and a glass case (d) $\Delta=0.25$  
for the 3D isotropic kicking with $\ka=10$ and $\ell=0.05$.
The corresponding surface for the Lindemann parameter $\ath_c(s,s')$ of the case of (c) is shown in (e).  
\label{fig2}}
\end{figure}

Besides the Eyink variational method, we also calculated
$\af$, $\ath$, and therefore $n_c$ by another method. 
From the mechanical feedback procedure, we first obtain 
the mapping from a hardsphere environment to an 
effective harmonic potential
$\ath=\ath(\af)$. Here $\ath$ depends on the steady-state probability
distribution of its neighbors.
Conversely $\af$ can be viewed as the final effective 
spring constant of a kicking particle in an 
harmonic potential of $\ath$. Next we numerically solve
$\af=\af(\ath)$ from single particle master equations using 
a variational method based on the 
square hermitianized
operator $\hat{l}^{\dagger}\hat{l}$  
with single particle trial functions.  
The two sets of operations are iterated to obtain a
pair of self-consistent results $(\af, \ath)$. 
The critical density predicted by this self-consistent squared hermitian
variational method agrees very well with results from the nonhermitian 
variational method.
The difference of instability density is less then 0.1\%  
when $\Delta <1$.
The corresponding critical $\af_c$ are also  similar. 
The two methods do give different results when $\Delta \gg 1$.

For persistent kicking, the trial functions have to be modified. Each
localized particle now has a distribution of locations of the form
$\psi^G(\vec{r}\;'=\vec{r}- \vec{b}; \overline{\overline{\alpha}})\sim
\exp(-\overline{\overline{a}}:\vec{r}\;'\vec{r}\;')$.
Here $\vec{b}$ is an off-center shift vector parallel to $\hat{n}$.
We must consider the effects 
of the variational parameter $b$ on $\ath$ along with the direct
changes of  $\overline{\overline{a}}$. The additional
decrease of $\ath$ arises  
from the distortion of the structures caused by always
kicking in the same directions.
We model the distortion effect of persistent kicking
on the pair distributions by
replacing each initial position with a dispersed distribution. For the crystal, 
this means the initial neighboring position of $\vec{R}_j$
is replaced by an average over positions $\vec{R}_j+ b \hat{n}$ with $\hat{n}$
is an arbitrary unit direction. Likewise, the radial distribution function of the glass
case is broadened from the initial $g_{b=0}(\vecr)$ to
$g_{b}(\vecr)= \int g_{b=0}(\vec{r'})({1\over 4 \pi b^2})\delta(|\vec{r'}- \vecr|-b) d \vec{r'}$.
In this case an additional normalization of the first peak enforces the condition
$g(r)=0$ for $r<1.$

For persistent kicking, $s$ has similar effects on $\overline{\overline{\alpha}}$ 
as were found for the isotropic case. The shift $b$ 
agrees quite well (within several percentages for the practical range of parameters)
with the estimate $\ka \ell \over 2 \ath D$ based on a small $\ell$ expansion
of the master equation. The resulting displacement amplitude $b$ is large enough to distort 
the stable structure 
so that any increased stability that may arise from kicking (if any) is now very
modest as shown in Fig.~3.

\begin{figure}
\includegraphics[scale=0.43, angle=-90]{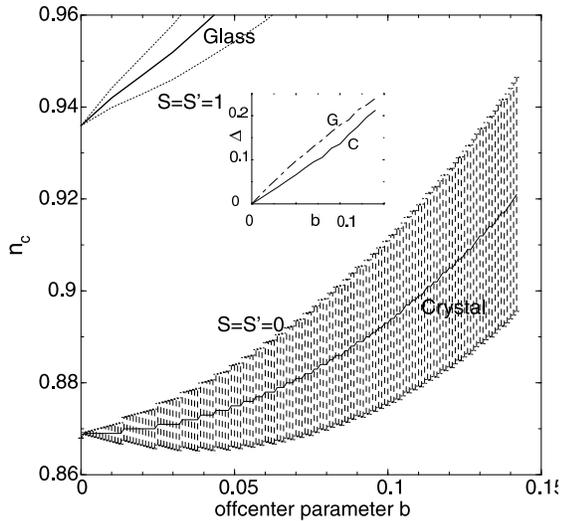}
\caption{The instability density of the persistent motorized fcc lattice
and the glass as parametric functions of $b$, which depends on $\Delta$. 
Here $\ell=0.05$. Both are bounded with
middle line $s+s'=1$, with upper and lower bound $s=s'=$ 0 and
1 respectively.
\label{fig3}}
\end{figure}

Since the kicking noise enlarges the  stability region 
in the isotropic case with $s+s'>1$, we wondered whether 
susceptible kicking may sometimes actually
decrease the effective temperature of this nonequilibrium system.
An effective temperature can be
defined by the fluctuation-dissipation relation 
even for far-from-equilibrium situations like the motorized 
crystal~\cite{CKP97}.  The ratio of the thermal temperature
to the  effective temperature is also called FDT violation factor.
The effective temperature can be computed by comparing the 
fluctuations of a motorized particle to its response to an external force.
$T^{eff}$ depends on the frequency or time duration, 
the absolute time (in case of an aging system)
and even possibly on the choice of the observable itself. 
To compute the needed time-dependent quantities, we solve the time dependent
master equations for
nonhermitian operators, again using a Gaussian ansatz 
characterized now with dynamic first and second moments. 

For illustration, we carry out the analysis of the dynamics for
the one dimensional symmetric case in a harmonic potential $\ath x^2$  with $s=s'$.
Thus $\hat{L}_{PF}~\psi= D \p^2_{xx} \psi - \p_x [\beta D f^h(x) \psi] $
and $\hat{L}_{NE}~\psi(x) = {\ka \over 2} \psi(x-\ell) e^{\eps(x-\ell,\ell)}+
  {\ka \over 2} \psi(x+\ell) e^{\eps(x+\ell, -\ell)}
- {\ka\over2} \psi(x) [e^{\eps(x,\ell)}+e^{\eps(x,-\ell)}]$
with $\eps(x, \ell)=  - 2 s  \ath x \ell- s \ath \ell^2.$
The parameters in the  Gaussian ansatz
$\psi(x, \mathbf{m}(t))= [2\pi \tilde{m}_2(t)]^{-1/2} \exp \lp-{[x-m_1(t)]^2\over
2 \tilde{m}_2(t)}\rp$
with $\tilde{m}_2:= m_2 -m_1^2$ and $m_i=\mean{x^i}_\psi$
satisfy the time-dependent dynamics described
by a set of differential equations with $\kap=\ka\exp(-s\ath \ell^2)$:
$\p_t{m}_1 = -2 D \ath m_1  - \kap \ell e^{2( s \ath \ell)^2 \tilde{m}_2}\sinh(2 s \ath \ell m_1)$
and
$\p_t{m}_2 = 2D - 4 D \ath \tilde{m}_2 + \kap \ell^2 e^{2(s \ath \ell)^2 \tilde{m}_2}\cosh(2 s \ath \ell m_1)
- 2 \kap \ell  e^{2 (s \ath \ell)^2 \tilde{m}_2} [m_1 \sinh(2 s \ath \ell m_1)
+ 2 s \ath \ell \tilde{m}_2 \cosh(2 s \ath \ell m_1)]$.

To obtain the Green's function
$G(x,x';t,0)$, $m_1(t)$ and $m_2(t)$  must satisfy the above equations with the initial conditions
$m_1(0)= x'$ and $m_2(0)= {x'}^2$.
We denote $\mathbf{m}(t;x')$ for this pair solution of the differential equation.
It is easy to see that for $\ka=0$, the exact expression of the Green's function
of an equilibrium system is  recovered.
\mycomments{At $t$ tends to $\infty$, the steady-state distribution has $ m_1^*=0$ and
$$m_2^* = 1/( 2\ath) { 1+ {1\over2} {\kap \ell^2 \over  D} \exp[{2 (s \ath \ell)^2} m_2^*]
\over  1+ s {\kap \ell^2 \over  D} \exp[{2(s \ath \ell)^2} m_2^*]}$$
The adamant motor ($s \to 0$) gives $m_2^* > 1/(2 \ath)$ while the
susceptible motor ($s \to 1$) always gives $m_2^* < 1/(2\ath)$. The behavior
changes at $s={1\over 2}$ just as for the
3D results.}
Green's function yields the correlations and responses.
The correlation functions are given by $C(t)= \int dx' x' m_1(t;x') \psi(x', \mathbf{m}^*)$. Here $*$
donates the steady-state value.
While the response to a pulse is $R(t)= \int dx' ( \beta D /{m_2^*}) x' m_1(t;x') \psi(x',\mathbf{m}^*)$.
Combining these yields the effective temperature
\begin{widetext}
\begin{equation}
 {T^{eff}(t)\over T^{th}} \equiv -\beta {\p_t C(t) \over R(t)} =
 \lp{ 1+ {1\over2} {\kap \ell^2 \over  D} \exp[{2 (s \ath \ell)^2} m_2^*]
\over  1+ s {\kap \ell^2 \over  D} \exp[{2 (s \ath \ell)^2} m_2^*]}\rp\times
\lp 1+{\kap \ell \over 2 D \ath}
{\int e^{2 {(s \ath \ell)^2} \tilde{m}_2(t;x')}\sinh(2 s \ath \ell m_1(t;x')) x' \psi^*(x') dx'
\over \int m_1(t;x') x' \psi^*(x') dx'} \rp 
\end{equation}
\end{widetext}

Thus we see that FDT violation
is a product of two ratios. One ratio is the steady-state
variance compared with the corresponding thermal equilibrium value.
The other ratio depends on the rate at which
the system reaches the steady state.
I.e., the larger the variance and the faster the
dynamics,  the hotter the system and vice versa.
Compared with the cases without kicking,
susceptible motors yield a smaller variance while
on the other hand, they relax faster.
In the short time limit, $m_1(t;x') \to x'$
and $\tilde{m}_2(t;x')\to 0$,
and the second ratio becomes
$ 1+ s {\kap \ell^2 \over D} e^{{2(s \ath \ell)^2} {m}^*_2}$.
For long times, $m_1\to 0$, $\sinh(s a \ell m_1) \to s a \ell m_1$,
and $\tilde{m}_2\to m^*_2$. We find
at this limit the value of the ratio
as $ 1+ s {\kap \ell^2 \over D} e^{{(s a \ell)^2\over 2} {m}^*_2}$,
exact same as the short time limit. Therefore
$T^{eff}(t=0)= T^{eff}(t=\infty)=
1+ {1\over 2}{\kap \ell^2 \over D} e^{{2(s \ath \ell)^2} {m}^*_2}$.
Yet for intermediate times, the ratio is not constant and differs from
either limiting value.
Generally, $T^{eff}>T^{th}$, i.e.,
the system is ``hotter''
although chemical noise apparently enlarges the stability range of the
localized phase.

We have studied the stability and dynamics of localized nonequilibrium
structures of motorized particles. 
The nonequilibrium noise from kicking motors sometimes
increases the effective spring constant and 
enlarges the mechanical stability range of both crystal and frozen glass
structures. We see that for systems like the cytoskeleton, nonequilibrium
noise may speed up the dynamics without sacrificing structural stability.
This model can be further developed to include anisotropy of the
particles or under other types of nonequilibrium noise or driven forces. We have also
studied the dynamics and effective temperatures~\cite{CKP97} of this system.
Besides taking this solid-state viewpoint, 
one can also study the transition from the liquid side
by mode coupling theory, a problem for future works.

This work is supported by NSF and CTBP.
T.S.~thanks Prof.~J.A. McCammon for his kind help and support,
especially in the early stage of the work.

\bibliographystyle{apsrev}

\end{document}